\documentclass[a4paper,11pt]{article}
\pdfoutput=1 % if your are submitting a pdflatex (i.e. if you have
\usepackage{jcappub} % for details on the use of the package, please
\usepackage[T1]{fontenc} % if needed

\usepackage{color} %
\usepackage{threeparttable}
\usepackage{url}

\newcommand{\mtrv}[1]{{\textcolor{black}{#1}}}
\newcommand{\ryrv}[1]{{\textcolor{black}{#1}}}
\newcommand{\rytrv}[1]{{\textcolor{black}{#1}}}

\title{\boldmath A Search for a Contribution from Axion-Like Particles to
the X-Ray Diffuse Background Utilizing the Earth's Magnetic Field}

\author[a,1]{R. Yamamoto,\note{Corresponding author. Present Address:
Advanced Institute for Science and Technology, 1-1-1, Umezono, Tsukuba,
Ibaraki, Japan}}
\author[a]{N.Y. Yamasaki,}
\author[a]{K. Mitsuda,}
\author[b]{and M. Takada}
\affiliation[a]{Institute of Space and Astronautical Science,\\ 3-1-1,
Yoshinodai, Chuo-ku, Sagamihara, Kanagawa, Japan}
\affiliation[b]{Kavli Institute for the Physics and Mathematics of the
Universe, University of Tokyo,\\ 5-1-5, Kashiwanoha, Kashiwa, Chiba, Japan}

\emailAdd{r.yamamoto@aist.go.jp}
\emailAdd{yamasaki@astro.isas.jaxa.jp}
\emailAdd{mitsuda@astro.isas.jaxa.jp}
\emailAdd{masahiro.takada@ipmu.jp}

\abstract{
\mtrv{The Axion Like Particle (ALP) is a hypothetical pseudo-scalar
particle beyond the Standard Model, with a compelling 
possible connection to dark matter and early universe physics. 
ALPs can be converted into photons via interactions with magnetic
fields in the universe, i.e., the so-called inverse Primakoff effect.
In this paper, we propose a novel method to explore ALP-induced photons
from X-ray data obtained from the {\it Suzaku} satellite,
arising from a possible interaction of ALPs with the direction-dependent
Earth's magnetic field viewed from the satellite.
{\it Suzaku} data is suitable for this purpose because its low-altitude
Earth orbit result in intrinsically low cosmic-ray background radiation.
We study whether the X-rayd diffuse background (XDB)
spectra estimated from the four deep fields collected over eight years,
vary with the integrated Earth's magnetic strength in the direction of
each target field at each observation epoch, which amounts to
$10^2$~Tm--a value greater than that achieved by terrestrial
experiments due to the large coherent length.
From the detailed analysis, we did not
find evidence of the XDB cofifdence level spectra having dependence on
the Earth's magnetic strength.
We obtained 99\% confidence level upper limit on a possible residual
contribution to the cosmic X-ray background (CXB) surface brightness to
be $1.6\times 10^{-9}~{\rm ergs~s}^{-1}{\rm cm}^{-2}{\rm sr}^{-1}$ 
normalized at $10^4$ T${}^2$ m${}^2$
in the 2--6~keV range, which
corresponds to 6--15\% of the observed CXB brightness, depending on which
model of unresolved point sources are used in the interpretation.
It is consistent with 80--90\% of the CXB now being resolved
into point sources.
}
}

\usepackage{version}	% required for `\comment' (yatex added)
\begin{document}
\maketitle
\flushbottom

\section{Introduction}
Various cosmological observations have provided strong evidence for the
existence of dark matter.
However, if dark matter is to be an elementary particle, it is a
yet-unknown particle beyond the Standard Model. 
The Axion-Like Particle (ALP) is a
\mtrv{hypothetical pseudo-scalar particle} beyond the Standard Model,
and is a consequence of the quantum field for conserving CP symmetry in
the strong interaction \citep{Peccei1977, Weinberg1978}.
ALPs are attractive because they act like cold dark matter (CDM) in the
formation of cosmic structure.
It is possible that ALPs are created by
a decay of other CDM-candidate particles.
If ALP mass is too low, a direct experiment with present
techniques is unlikely to find it.
A possible channel is to observe photons, which are created by ALPs
following the inverse Primakoff process in an electromagnetic field.
As we show in detail in Section 2, 
the ALP-photon conversion probability $P_{a \rightarrow \gamma}$ is
approximately proportional to the squared product of
magnetic field \mtrv{strength} orthogonal to the ALP 
\mtrv{momentum direction,}
$B_\perp$, and the length, 
$L$, i.e., $P_{a \rightarrow \gamma}$ $\propto$
$\left( B_\perp L \right)^2$.

\mtrv{There have been many attempts to detect ALP signals in terrestrial
experiments and astronomical observations. One candidate signal is
in the direction of galaxies or galaxy clusters, which
was proposed to be due to 
ALP interaction with inter-stellar or galactic magnetic fields}
\cite{Cicoli2014, Conlon2014, Conlon2015a}, \mtrv{although the results
are still under discussion.
If ALPs are CDM itself or produced from a decay of CDM at cosmological
distances and if those can be observed, the distribution of ALPs or
ALP-induced photons via its interaction with magnetic fields in cosmic
structures should appear to be isotropic in the sky to the zero-th order
approximation, unless we have high-sensitivity and high-angular
resolution data to resolve the distribution tracing inhomogeneous cosmic
structures in the universe.}

\mtrv{In this paper, we propose a novel method to search for ALP-induced
photons from the satellite X-ray data, arising from the Primakoff
interaction of the ALPs with the Earth's magnetic field. 
The Earth's magnetic field is known to have a dipole structure around
the Earth in a north-south direction and we have a good knowledge of its
strength and field configuration from various observations.
Therefore, we can expect the ALP-induced photons in X-ray wavelengths,
if produced, to vary with the Earth's magnetic field strength integrated
along the line-of-sight direction for each observation, even if ALPs
arriving on Earth have an isotropic distribution in the sky.
To search for such ALP-origin X-ray radiation, we focus on the
{\it Suzaku} X-ray data in the four deep fields, collected over
eight years. These fields had been observed frequently by the {\it Suzaku},
but the magnetic field strengths varied with each observation
depending on its location in the orbit. 
For a null hypothesis of ALP-induced photons, the diffuse X-ray
background brightness estimated from the same field should {\it not}
show any dependence on the integrated magnetic field.
This is the signal we will search for in this paper.
{\it Suzaku} data is suitable for our purpose, because the
satellite, compared to the
{\it XMM-Newton}\footnote{\url{http://sci.esa.int/xmm-newton/}} or
{\it Chandra}\footnote{\url{https://chandra.harvard.edu/}}, has lower
background due to its low-altitude orbit that prevents cosmic rays from
penetrating the X-ray detectors \cite{Mitsuda2007}.}

\mtrv{Our study is somewhat similar to}
Fraser et al., (2014) \cite{Fraser2014}, which claimed a detection of
seasonal variations in the {\it XMM-Newton} X-ray data.
\mtrv{The work claimed that}
the X-ray flux modulation
\mtrv{at a level}
4.6 $\times$ 10$^{-12}$ ergs s$^{-1}$ cm$^{-2}$
deg$^{-2}$ in 2--6 keV 
\mtrv{might be due to a conversion of solar axions by their interaction
with the Earth's magnetic field
\citep[also see][]{Davoudiasl2006, Davoudiasl2008}.}
However, Roncadelli and Tavecchio, (2015) \cite{Roncadelli2015} claimed 
that the {\it XMM-Newton} satellite which never points toward the Sun
cannot observe
\mtrv{such ALP-induced photons originating from solar axions due to
momentum conservation.}

\mtrv{The structure of this paper is as follows. In
Section~\ref{sec:basics}, we briefly review basics of the inverse
Primakoff effect and how photons can be induced by
the effect from ALPs that are created by dark matter in the
expanding universe. In Section~\ref{165855_4May19} we show the main
results of this paper using the {\it Suzaku} data, when combined with
data of the Earth's magnetic field at each orbit of the {\it Suzaku}
satellite at each observation. 
Section~\ref{sec:discussion} cotains the discussion and conclusion.}

\section{Process of photon emission from ALPs}
\label{sec:basics}
\mtrv{In this section we describe a mechanism of photon emission
from ALPs via the interaction with magnetic fields.
To do this, we consider a model in which dark matter, which fills up
space of the universe, preferentially decays into ALPs.
This model is an example of moduli dark matter model in a
string-theory-inspired scenario.}

\mtrv{When a dark matter particle decays into two ALPs, i.e. }
DM $\rightarrow$ 2ALPs,
\mtrv{each ALP has}
a monochromatic energy $E_a = m_\phi /2$,
where $m_\phi$ is the mass of the dark matter \ryrv{particle}. 
The emissivity of
\mtrv{DM $\rightarrow$ 2ALPs decaying process}
is
\mtrv{given in terms of}
the energy density of dark matter,
$\rho_\phi \left( r\right)$, 
the decay rate, $\Gamma_{\phi \rightarrow 2a}$, and $m_\phi$ as
\begin{equation}
 \epsilon_a = \frac{2 \rho_\phi \left( r\right)
 \Gamma_{\phi \rightarrow 2a} }{m_\phi}.
\end{equation}
Considering the spatial distribution of dark matter along the
line-of-sight direction, the ALP intensity, 
$I_{a,{\rm line}}$ [counts s$^{-1}$ cm$^{-2}$ sr$^{-1}$],
is \mtrv{given as}
\begin{equation}
 I_{a,{\rm line}} = \int_{\rm l.o.s.} \frac{2 \Gamma_{\phi \rightarrow
  2a}} {4 \pi m_\phi} \rho_\phi \left( r\right)~dr
 = \frac{S_\phi \Gamma_{\phi \rightarrow 2a}}{2\pi m_\phi},
\label{eq:line}
\end{equation}
\mtrv{at $E_a=m_\phi/2$, and $S_\phi$ is the column density of dark
matter in the line-of-sight direction \citep{Sekiya2016}, defined as}
\begin{equation}
 S_\phi = \int_{\rm l.o.s.} \rho_\phi (r)~dr.
\end{equation}
In this case, the converted photon spectrum is a line emission.

\mtrv{If dark matter is uniformly distributed in the universe, we
would observe a continuum spectrum of the ALP intensity
because free-streaming ALPs undergo a cosmological redshift in
the expanding universe.
Assuming light-mass ALPs, i.e. relativistic ALPs, produced by
dark matter decay, a superposition of line spectra over 
different redshifts leads us to observe a continuum spectrum of ALPs
\cite{Kawasaki1997,Asaka1998}: }

\begin{eqnarray}
 \frac{dN}{dE_a} &=& \int_{\rm l.o.s.} \mathrm{d}r~
 \frac{\Gamma_{\phi \rightarrow 2a}}{4 \pi m_\phi}
 \rho_\phi \left( r\right) \times 2
 \delta_D\! \left( E_a \left( 1+z \right) - m_\phi /2 \right) \\
 &=&
 \frac{\sqrt{2} c \Gamma_{\phi \rightarrow 2a} \rho_{\phi_0}}{\pi H_0}
 ~m_\phi^{-\frac{5}{2}} ~E_a^{\frac{1}{2}}
 ~f\left( \frac{m_\phi}{2E_a} \right)
 \label{eq:con}
\end{eqnarray}
\mtrv{where $\delta_D(x)$ is the Dirac delta function, and the function
$f(x)$ is defined as}
\begin{equation}
 f(x) \equiv \left\{  \Omega_{m0} +
 \left( 1-\Omega_{m0} -\Omega_{\Lambda 0}
 \right)/x - \Omega_{\Lambda 0}/x^3 \right\}^{-\frac{1}{2}}.
\label{124611_31May18}
\end{equation}
\mtrv{In the above equation}
$z$ is the redshift at decay, $\rho_{\phi_0}$ is the present energy density,
$H_0$ is present the Hubble constant, $\Omega_{m0}$ and $\Omega_{\Lambda 0}$
are the density parameters of 
\mtrv{non-relativistic}
matter and the cosmological constant, respectively.
The spectral shape of ALPs is transcribed as a simple power-law function 
whose number index is a positive value of $+1/2$.
In this case, the converted photon spectrum is also expected as a power-law 
function with a photon index of $+1/2$.

The ALP-photon conversion probability in a vacuum with a magnetic field
via inverse Primakoff effect is given 
in Ref.~\cite{VanBibber1989} as
\begin{equation}
 P_{a \rightarrow \gamma} \left(x\right)
 = \left| \frac{g_{a \gamma \gamma}}{2} \int_0^{x} B_\perp
 \left( x'\right)
 \exp \left( -i \frac{m_a^2}{2E_a} x' \right) dx' \right|^2,
\label{193602_11Jun16}
\end{equation}
with
\begin{equation}
 B_\perp \left(x'\right) \equiv \left|
 \vec{B} \left(x'\right) \times \vec{e}_a
 \right|.
\end{equation}
Here,  $g_{a \gamma \gamma}$ is an ALP-photon
coupling constant, $m_{a}$ and $E_{a}$ are mass and energy scales of ALP, and
$B_\perp(x)$ is the perpendicular component of magnetic field to the 
ALP \mtrv{momentum direction, denoted as} $\vec{e}_a$.
The ALP-photon momentum transfer $q$ is defined as
\begin{equation}
 q = \frac{m_a^2}{2E_a}. \label{181820_20Jun16}
\end{equation}
Assuming the $B_\perp (x')$ is uniform in the range $0<x'<L$,
we can write Equation \eqref{193602_11Jun16} as:
\begin{equation}
 P_{a \rightarrow \gamma} = \left( \frac{g_{a\gamma \gamma} B_\perp}{2}
			    \right)^2~2L^2~
 \frac{1- \cos \left( qL \right) }{(qL)^2}.
 \label{163650_19Feb16}
\end{equation}
In the limit of 
\mtrv{light}
ALP masses \mtrv{compared to the photon energy scale satisfying $q L\ll 1$,}
$1- \cos \left( qL \right) \simeq \left( qL \right)^2 /2$, and
the conversion rate is simply given by
\begin{equation}
 P_{a \rightarrow \gamma} = \left( \frac{g_{a \gamma \gamma} B_\perp
			     L}{2} \right)^2.
\label{145821_20Feb16}
\end{equation}
under the coherence condition of
\begin{equation}
 qL < \pi ~ \rightarrow ~ m_a < \sqrt{\frac{2\pi E_a}{L}}
\label{qcon}
\end{equation}
\mtrv{The following analysis uses
Equation~\eqref{163650_19Feb16} to constrain the ALP-photon coupling
constant.}

\mtrv{As shown above,}
the probability of ALP \mtrv{particles converting to photons}
proportional to $(B_\perp L)^2$ in the light mass limit.
\mtrv{Plugging typical values of the strength and coherent length scale
of Earth's magnetic field,}
Equation~\eqref{145821_20Feb16} gives
\begin{eqnarray}
 P_{a \rightarrow \gamma} &\simeq & 2.45 \times 10^{-21}~
 \left( \frac{g_{a \gamma \gamma} }{10^{-10}~{\rm GeV^{-1}}} \right)^2
 \left( \frac{B_\perp L}{{\rm T~m}} \right)^2
 \label{con_g10}
\end{eqnarray}

\section{Analysis and results: A search for the correlation between
 residual {\it Suzaku} background radiation with the Earth's magnetic
 strength}
\label{165855_4May19}
\subsection{Selection of blank sky observations from {\it Suzaku}
  archival data} 
\label{subsec:field_selection}

\begin{figure}[htbp]
\centering % \begin{center}/\end{center} takes some additional vertical space
\includegraphics[width=.48\textwidth,bb=0 0 483 480]{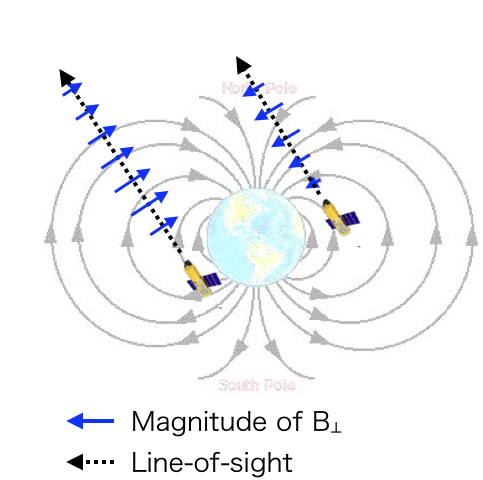}
\hfill
\includegraphics[width=.48\textwidth,bb=0 0 640 480]{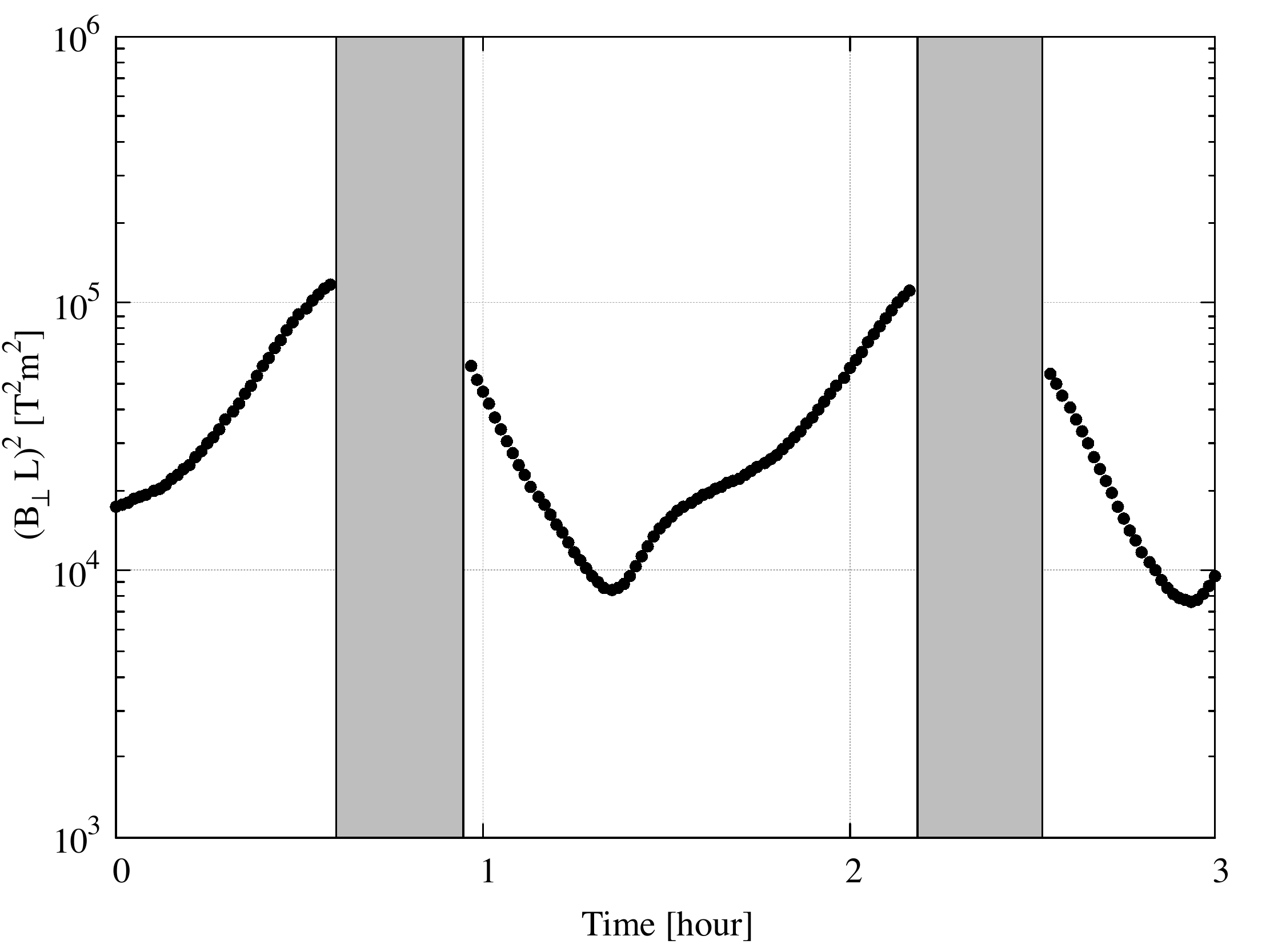}
\caption{Left: Schematic view of  the  position and
 observation direction of {\it Suzaku} satellite relative to the Earth's
 magnetosphere. Right: Time dependence of $\left( B_\perp L\right)^2$ in
a Lockman hole observation. 
Gray hatched regions show periods of the Earth occultation, i.e. 
 the Earth exists between a target and {\it Suzaku}.}
\label{112957_31May18}
\end{figure}

\mtrv{To locate ALP-induced photons, we use the {\it Suzaku} X-ray data,
and search for photons in the detector's field of view (FoV)
depending on the integrated magnetic strength along the line-of-sight
direction, $\left(B_\perp L\right)^2$.
Because most X-ray data contains X-ray emission photons from targeted
or unresolved sources, we need to study the 
X-ray diffuse background (XDB) in blank fields, and search for a residual
signal in the background that is
correlated with the magnetic strengths following the scaling of
$(B_\perp L)^2$.} The X-ray satellite {\it Suzaku} is suitable for 
\mtrv{this study because of its}
low instrumental background noise and 
\mtrv{the low background radiation from cosmic rays
(compared to other X-ray satellites) due to its low-altitude Earth orbit;}
an altitude of $\sim$ 570 km and an inclination of $31^\circ$
\mtrv{from the Earth's equatorial plane, where the Earth's magnetic field
prevents cosmic rays from penetrating the satellite's detectors} 
\cite{Mitsuda2007}.
Figure \ref{112957_31May18} is a schematic \mtrv{illustration}
of the {\it Suzaku} orbit 
\mtrv{and the Earth's magnetic field configuration. Even if the
satellite observes the same field or the same angular direction--as
denoted by the black dotted line--the integrated strength of
perpendicular magnetic components along the line-of-sight direction
varies with the satellite position.
The {\it Suzaku} satellite orbits the Earth with a period of
approximately 96 minutes and it causes a
modulation of the integrated magnetic strength $(B_\perp L)^2$ with the
orbit, or when the target field is observed.
Thus, we expect variations in
the ALP-induced photons, if they exist, depending on the strength
$(B_\perp L)^2$.}
We calculated \mtrv{the Earth's magnetic field every 60 seconds for
each line-of-sight direction of a given target field using the
software,}
{\it International Geomagnetic Reference Field:} the 12th generation
(IGRF-12 \cite{Thebault2015}) for up to 6 times the Earth's radius
($R_E$), where \mtrv{typically} $B$ $\sim$ $10^{-7} {~\rm T}$.
The right panel in Figure \ref{112957_31May18}
\mtrv{shows a typical case of $(B_\perp L)^2$ as a function of the
satellite position or equivalently the observation time.}
\mtrv{It can be found} that a typical 
value of $\left( B_\perp L\right)^2$ is of order of
$10^{4}$--$10^{5}$ T$^{2}$m$^{2}$,
\mtrv{which is greater than that of terrestrial experiments such as the
CAST experiment\footnote{\url{http://cast.web.cern.ch/CAST/CAST.php}}.}
If we apply the non-oscillation condition 
\mtrv{of $qL \ll 1$}
(Equation~\eqref{qcon}), the
corresponding mass of ALP is limited to be $m_a$ $\leq$ $\mu$eV if
\mtrv{we assume that the}
converted photons are in X-ray \mtrv{wavelengths}.
\mtrv{Note that we considered an oscillation regime of $q L\sim 1$ to
obtain constraints on the ALP-photon coupling constant.}

\mtrv{To estimate the XDB spectrum, we
consider blank sky data from four deep fields selected from}
the {\it Suzaku} archives as tabulated in Table \ref{130850_22Jun16}.
The selection criteria are as follows.
\begin{enumerate}
 \item No bright sources \mtrv{in the FoV of {\it Suzaku} X-ray Imaging
       Spectrometer (XIS) \cite{Koyama2007}, and compact sources in the
       FoV are already identified and can be masked in our analysis.}
 \item Galactic latitudes of $|b| > 20^\circ$ to avoid
       X-ray emission \mtrv{originating from sources in}
       the Galactic disk \citep{Masui2009}.
 \item \mtrv{Sufficiently distant from regions of high X-ray diffuse
       emissions such as the North Polar Spur.}
 \item Exposure time obtained by standard processing \mtrv{should be}
       more than 200~ksec.
\end{enumerate}
\mtrv{The above criteria are met by the following four fields, also shown
in Table~\ref{130850_22Jun16}. First, we use the multiple observation data in}
the Lockman hole field, which is a famous
\mtrv{region with minimum neutral hydrogen column density}
that was annually observed with {\it Suzaku} for calibration.
\mtrv{We also use the data in the} 
South Ecliptic Pole (SEP) and North Ecliptic Pole (NEP) fields.
\mtrv{Finally, we use the data in the field of high latitude, the neutral
hydrogen cloud or the so-called} MBM16 field. 

\begin{table}[htbp!]
\begin{center}
\begin{threeparttable}
\caption{Observation of long exposure background observation by {\it
 Suzaku} satellite}
\label{130850_22Jun16}
\begin{tabular}{l c c c c c c} \hline \hline
Field name & $\left( \alpha_{2000}, \delta_{2000}\right)$ & Num. of &
Total & Num. of & Exposure used & Obs. Year\\
~ & ~ & Obs. & exposure${}^\ast$ & events${}^\dagger$ &
		     in this analysis${}^\dagger$ & ~ \\
~ & ~ & ~ & [ksec] & [counts] & [ksec] & ~ \\ \hline
Lockman hole & (162.9, 57.3) & 11 & 542.5 & 5595 & 210.7 & 2006--2014 \\
MBM16 & (49.8, 11.7) & 6 & 446.9 & 10755 & 231.8 & 2012--2015 \\
NEP & (279.1, 66.6) & 4 & 205.0 & 7666 & 221.9 & 2009\\ 
SEP & (90.0, -66.6) & 4 & 204.2 & 6102 & 180.2 & 2009\\ \hline
\end{tabular}
\begin{tablenotes}[flushlef]
\begin{footnotesize}
\item[$\ast$] \ryrv{Exposure time at each XIS after the standard data
 processing pipeline.}
\item[$\dagger$] \ryrv{The sum of the three XIS exposure time after
 extra data reduction and $(B_\perp L)^2$ selection, used values in this
 paper.}
\end{footnotesize}
\end{tablenotes}
\end{threeparttable}
\end{center}
\end{table}

\begin{figure}[htbp!]
\begin{center}
\includegraphics[scale=0.35]{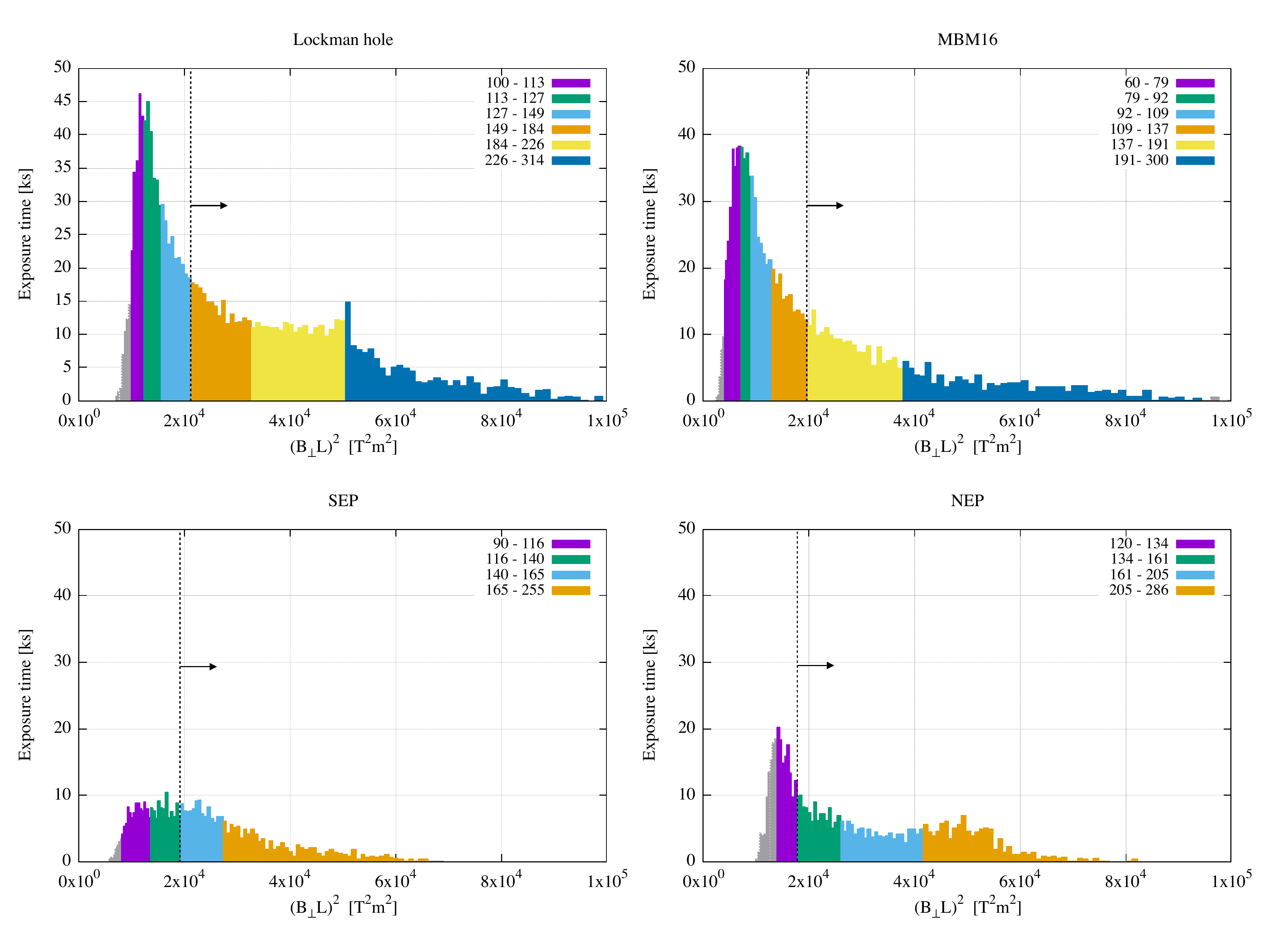}
\caption{The histograms of $\left( B_\perp L \right)^2$ during
 observation of 4 direction. Binning of exposure time, to 
 obtain an almost equal number of photons in each class of
 $\left(B_\perp L \right)^2$ are shown.
 Note that only $\left(B_\perp L \right)^2$ $\geq$ 2$\times10^{4}$
 T$^{2}$ m$^{2}$ are used for spectral analysis as shown in
 section 3.2.}
\label{153849_24Jun16}
\end{center}
\end{figure}

\mtrv{We use the data reduction pipelines, the {\it Ftools} in HEAsoft
version 6.16 and XSPEC version 12.8.2, to analyze the X-ray data in the
four fields of Table~\ref{130850_22Jun16}, collected from the archive of
{\it Suzaku} XIS}.
\mtrv{To avoid a possible contamination from high X-ray background,}
we removed data during the South Atlantic Anomaly region, 
Earth occultation, low elevation angle from the Earth's rim, and low
cut-off-rigidity (COR; < 8 GV/c) regions.
We \mtrv{stacked} the X-ray image in the 0.5--7 keV band 
\mtrv{for each of the four fields, where we}
removed the point sources whose flux is larger than 1 $\times$ $10^{-14}$
ergs s$^{-1}$ cm$^{-2}$, with a radius of 1.5 arcminutes
corresponding to encircled power function of 90\% for {\it Suzaku}'s
mirror.
We then calculated the $\left( B_\perp L \right)^2$ every
60 seconds \mtrv{for each observation, as a function of}
the satellite position in orbit and observing line-of-sight direction.
Figure \ref{153849_24Jun16} shows
the distribution of $\left( B_\perp L \right)^2$ in each
\mtrv{of the four fields}.
\mtrv{We subdivided the data into four to six bins of the
$\left(B_\perp L \right)^2$ values, where the binning was determined so
that each bin had almost the same photon statistics, as denoted by
the different colored histograms in the figure.}

\subsection{An assessment of non-Xray background contamination}
\label{sec:non_xray_background}

\mtrv{Before presenting the main results, we need to assess the level
of non-Xray background (NXB) contamination in the data.}
Although the NXB of {\it Suzaku} is \mtrv{usually}
low, 
\mtrv{there could be a residual NXB contamination, up to 16--50\% of the
observed CXB in the 2--6~keV range.}
\mtrv{A part of NXB is due to} fluorescence lines by materials such as
Si, Al, Au, and Ni around the detector.
\mtrv{These lines are distinguishable, if their emission lines are
identified at their corresponding energy scales, in each XIS spectrum,
as} identified in Ref. \cite{Yamaguchi2006}.
Not only X-rays but charged particles too can produce pseudo-events in
the CCD instrument.
Pseudo events produced by particles originating from cosmic rays in
orbit were studied by GEANT4 Monte-Carlo simulation.
The continuum spectra by pseudo events are
reproducible at an accuracy of 20\% in its amplitude in each energy
bin \cite{Murakami2006}.
The production processes of pseudo events are well understood, but the
input cosmic-ray flux varies by time and position of the satellite. 
The reproduction of the events was studied by using the event database
collected during the periods when the FoV was blocked by the night side of
the Earth (NTE) \cite{Tawa2008}.
It was found that the intensity of the background could be estimated as
a function of COR, and that the spectra were similar.
They proposed a background estimation method to make a spectrum from
$\pm$ 150 days of stacked night-Earth data weighted to reproduce the
distribution of COR.
They also found that the fluctuation of the background is larger than the 
simple Poisson statistics.
Uncertainty of reproducibility for a typical 50 ksec
exposure was reported to be 3.4\%, although the expected statistical
error by Poisson statistics is 1/10 \cite{Tawa2008}. 
This procedure is used as a standard background estimation for
{\it Suzaku} and adapted as a HEASoft tool.

In our analysis, we needed to sort the data by
$\left( B_\perp L \right)^2$, which correlated with the COR. If the orbital
position or $\left( B_\perp L \right)^2$ is a potential control parameter
of the NXB, it will affect our determination of the $\left( B_\perp
L\right)^2$ modulated signal.
COR parameters used in the {\it Suzaku} analysis (defined as COR2 in the
calibration database) are defined by the projected geographic
coordinates, and calculated by the geomagnetic model on 2006 Apr. 
Actual COR would change gradually with time, and the cosmic ray flux is
affected by the solar activity.
We thus stepped into further NXB analysis of  {\it Suzaku}, to evaluate
possible range of background fluctuation, and to define further data
reduction methods if needed.

\begin{figure}[htbp!]
\begin{center}
\includegraphics[scale=0.5,bb=0 0 640 480]{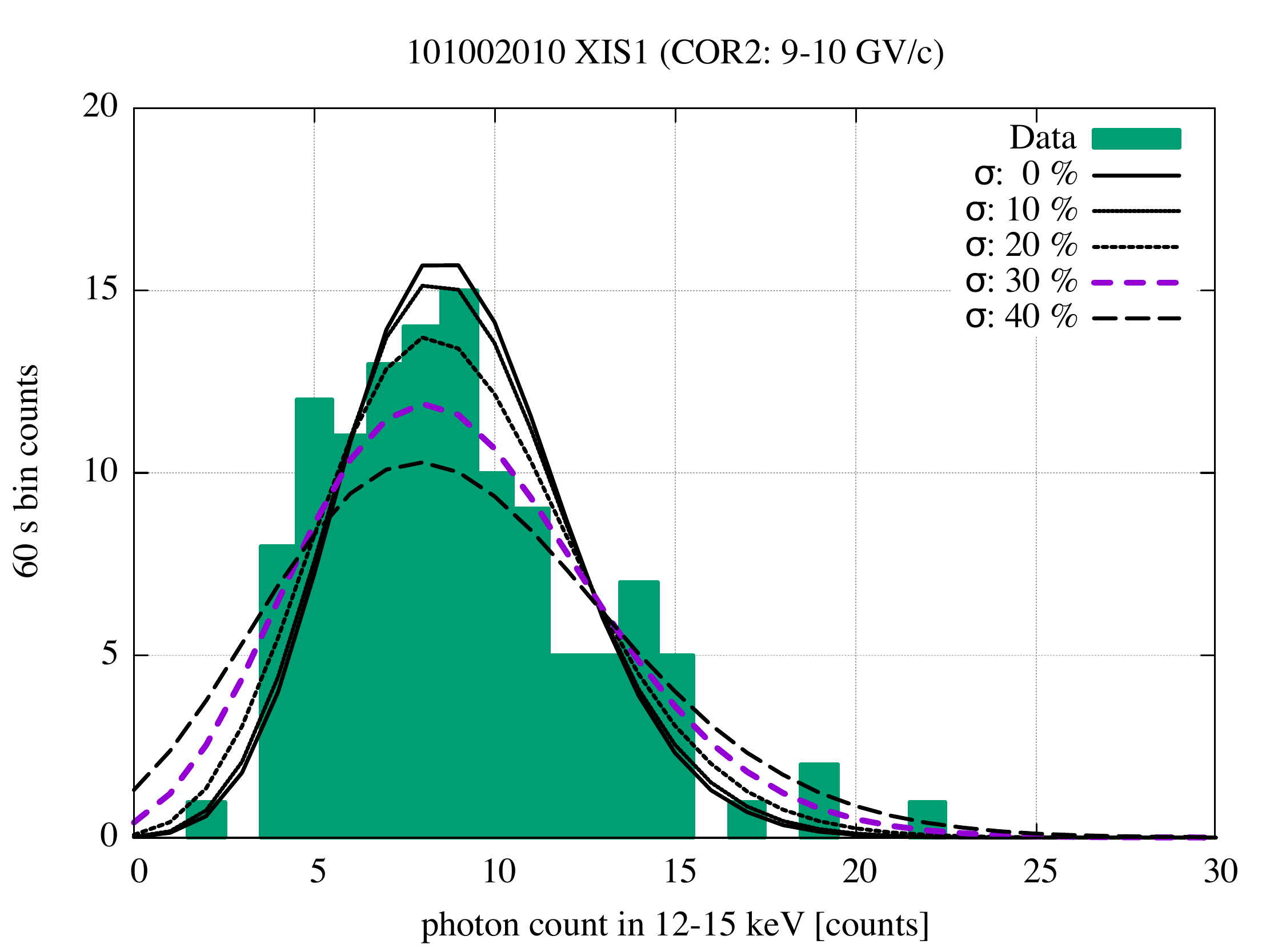}
\caption{A sample of histogram for observed events in 12--15 keV of
 XIS1 in COR range of 9 to 10 GV\slash $c$ and
 the probability distribution of each $\sigma \slash \mu$
 estimated from Equation \eqref{122042_23Jan17}.}
\label{calc_sample_short_term}
\end{center}
\end{figure}

We evaluated the fluctuation of input cosmic-ray flux by event
rates at 12--15 keV of the XIS1. As the effective area of
the X-ray mirror dropped rapidly above the Au L-edge below 1\%, the
event rate above 12 keV is considered to be an indicator of the cosmic
ray flux.
Due to the back-illumination structure, the background rate of XIS1 is
higher than the other  front-illuminated CCD, XIS0 and XIS3, and is more
sensitive. 
\ryrv{Apparently, the fluctuation of the background count rate exceeds
the Poisson statistics.
In \cite{Tawa2008}, the intrinsic fluctuation is evaluated as
$\sqrt{\sigma_{\rm calc}^{2}-\sigma_{\rm Poisson}^{2}}$.
We evaluate the intrinsic fluctuation as follows:
\begin{enumerate}
 \item Counting the number of events in 12--15 keV during the 60 sec
       for each COR range.
 \item Calculating an mean of the count every 60 sec bins, denoted as
       $\mu$, from the distribution of count rate as shown in histogram
       in Figure \ref{calc_sample_short_term}.
 \item Assuming a certain value $\sigma$, and simulating a $P_{\rm
       NXB}$ by Monte-Carlo method as shown in Equation
       \eqref{122042_23Jan17} (lines in Figure
       \ref{calc_sample_short_term}).
\begin{equation}
 P_{\rm NXB}\left( X=k \right) =
 \frac{\lambda\left( k, \sigma \right)^k e^{-\lambda\left( k, \sigma
 \right)}}{k!}, \quad \lambda \left( k, \sigma \right)
 = \frac{1}{\sqrt{2\pi \sigma^2}}
 \exp \left(-\frac{\left( k-\mu \right)^2}{ 2\sigma^2} \right),
\label{122042_23Jan17}
\end{equation}
where $k$ is an observation frequency per interval,
$\lambda \left( k, \sigma \right)$ is an average number of events per
interval, $\sigma $ is an estimated systematic error by the variation
of $\lambda \left( k, \sigma \right)$,
and $\mu$ is an average of observed NXB events.
 \item Comparing the observed and simulated histograms by Pearson's
       chi-squared test and obtaining 95\% upper limit for $\sigma$.
\end{enumerate}
It assumes that the variations of the mean value of the count rates
follow a Gaussian distribution,
and the detected count follows a Poisson distribution.
Thus, the observed count rate is expressed by a convolution of these
functions.
We use Pearson's chi-squared test to set a quantitative upper limit
for the short term variability. A sample of these tests is shown in
Figure \ref{calc_sample_short_term}.
The 95\% confidence range for the standard deviation of the Gaussian
is obtained as 22--39\% of the mean value.}

The background rate anomaly of the geographical position is checked
as follows.
We divided the orbital position projected onto the Earth's surface by
every $10^\circ$ in longitude
and $5^\circ$ in latitude as defined as {\tt Loc\_ID}, and sorted them
into 4 COR classes.
For each observation, the NTE count rate in 2--5.6 keV for
$\pm$150 days, which is used by the standard background estimation, was
accumulated on every {\tt Loc\_ID}. If the count rate at an 
{\tt Loc\_ID} is higher than the averaged value over the same COR class by
3$\sigma$, the events occurred at that {\tt Loc\_ID} were discarded from the
spectral analysis.

After these data were reduced, the count rate in 2--6 keV before and after
the standard background subtraction were plotted as a function of
$\left(B_\perp L\right)^2$. We found that there is an negative
correlation between the count rate and the $\left(B_\perp L\right)^2$,
contrary to the ALP origin signal prediction. 
We checked the count rate of the upper discriminator (PIN-UD) of the
Hard X-ray Detector (HXD)
%detector
onboard {\it Suzaku},
which corresponds to energy deposited by protons approximatelly $>100 $
MeV \cite{Kokubun2007},and found the same trend.
The PIN-UD is affected by the radio-activation of HXD itself, and 
cannot be used to estimate the XIS background.
We evaluated the correlation by a linear function fit, and decided that 
only those data satisfying $\left(B_\perp L\right)^2$ $\geq$ 2 $\times$
$10^4$ T$^2$ m$^2$ would be used for the analysis.

\subsection{Spectral analysis for $\left( B_\perp L \right)^2$ sorted data}
In the spectral analysis, we assumed that celestial diffuse emission of
each blank field is expressed by the sum of Cosmic X-ray Background
(CXB), Milky Way Halo (MWH) emission, Solar Wind Charge eXchange (SWCX),
Local Hot Bubble (LHB), and unknown
High Temperature Component (HTC) as studied by previous works
\cite{Yoshino2009,Sekiya2016,Nakashima2018}.
\ryrv{These are collectively called XDB.}
The surface brightness and
spectral parameters for the celestial emission can be varied by the FoV
in a reasonable range.
The ALP signal has a power-law spectral shape with a photon index of
$+1/2$, and with intensities proportional to $\left( B_\perp L \right)^2$. 
The NXB for each observation can be estimated by the standard background
estimation method
\cite{Tawa2008}, but the intensities can also be varied within the
fluctuation studied in the previous subsection.

Steps of spectral analysis for one observing direction are as follows;
\begin{enumerate}
\addtocounter{enumi}{-1}
 \item Apply standard data reduction for XIS 0,1,3 of each observation ID
        (a unit of archival data, events from continuous pointing for
       the same observation direction), point source removal, {\tt Loc\_ID}
       selection and $\left(B_\perp L\right)^2$ cut. 
       Response matrices \cite{Ishisaki2007} and template NXB by standard 
       method \cite{Tawa2008} are also prepared. 
       \label{163543_28May18}
 
 \item Accumulate the energy spectra in the 0.7--7.0 keV range for each
       XIS 0,3 and the 0.7--5.0 keV range for XIS 1
       subtract the standard NXB, and fit them simultaneously by an
       empirical X-ray background model, \ryrv{obtained the best-fit
       values and errors} with $\chi^2$ statistics
       and $C$-statistics \cite{Cash1979} in {\it Xspec}, and evaluate
       the validity of parameters. \label{162401_28May18}

 \item Divide the energy spectra by $\left(B_\perp L\right)^2$ values and
       fit them again simultaneously with $C$-statistics because of low
       photon statistics in each range. Check the consistency of
       spectral parameters obtained in step \ref{162401_28May18}.
       \label{164512_28May18}
       
 \item Add ALP emission model as a power-law function with a photon
       index of $+1/2$ and with a surface brightness proportional to the
       $\left( B_\perp L\right)^2$, and treat the background as
       a spectral model whose intensities can be tuned.
       \label{164529_28May18}
\end{enumerate}

The fitting model describing the diffuse X-ray emission is similar to that
used in \cite{Sekiya2016}; it is shown by 
\begin{displaymath}
apec_{\rm SWCX+LHB}+phabs(apec_{\rm MWH}+power{\mathchar`-}law_{\rm
 CXB}+apec_{\rm HTC}).
\end{displaymath}
The APEC (Astrophysical Plasma Emission Code) \cite{Smith2001}
\footnote{latest version is available at http://www.atomdb.org} is
an emission model from collisional equilibrium and optically thin plasma
installed in {\it Xspec} and applied to estimate the SWCX and LHB
blend, MWH, and HTC. 
The temperature of {\it apec} in the SWCX and LHB blend was fixed to
$kT =$ 0.1 keV \cite{Yoshitake2013}.
The typical temperature of the MWH is $kT =$ 0.15--0.35 keV
\cite{Yoshino2009,Sekiya2016},  
a part of the blank sky spectra requires a HTC with
$kT=0.6\mathchar`-0.9$ keV
to describe emission of approximately 0.9 keV \cite{Sekiya2016}. 
The CXB was represented by a power-law emission model with a photon
index of $\sim 1.4$. 
The solar abundance table of {\it apec} model was given by \cite{Anders1989}.
{\it Phabs} describes the absorption by the Galactic interstellar medium,
whose column density is
fixed from the LAB(Leiden/Argentine/Bonn) survey \cite{Kalberla2005} database. 
Steps \ref{163543_28May18} --\ref{162401_28May18} are the standard spectral
fitting procedure for {\it Suzaku}, 
and the parameters obtained in Step \ref{162401_28May18} were consistent
with each other within 
90\% error, and with previous works like 
Sekiya et al. (2016) \cite{Sekiya2016}. 

In step \ref{164512_28May18}, we divided the data by $\left( B_\perp L \right)^2$.
For example, in the case of Lockman hole, there were 11 observations, sorted by
$\left( B_\perp L \right)^2$ into 3 classes, and 3 CCDs,
thus 99 spectra were fitted simultaneously with the same emission parameter.
The number of the energy spectra for Lockman hole,
MBM 16, SEP, and NEP, are 99, 36, 24, and 36, respectively.
The degrees of freedom in the spectral fit also increased, and the number of
photons in each energy bin decreased.
We applied $C$-statistics, which assumes that the data
follows a Poisson distribution and uses the likelihood ratio to be
minimized, and confirmed that the obtained parameters are consistent
with Step \ref{162401_28May18}.
Because we would divide the spectra in later analysis, complex structure
(mainly of Oxygen lines below 0.7 keV) are not well resolved. We only
used the data in from the 0.7--7.0 keV range.
Some components whose
intensities were consistent with null were ignored by setting the
intensities to 0. 

\begin{figure}[htbp!]
\begin{center}
\includegraphics[scale=0.35,bb=0 0 1024 768]{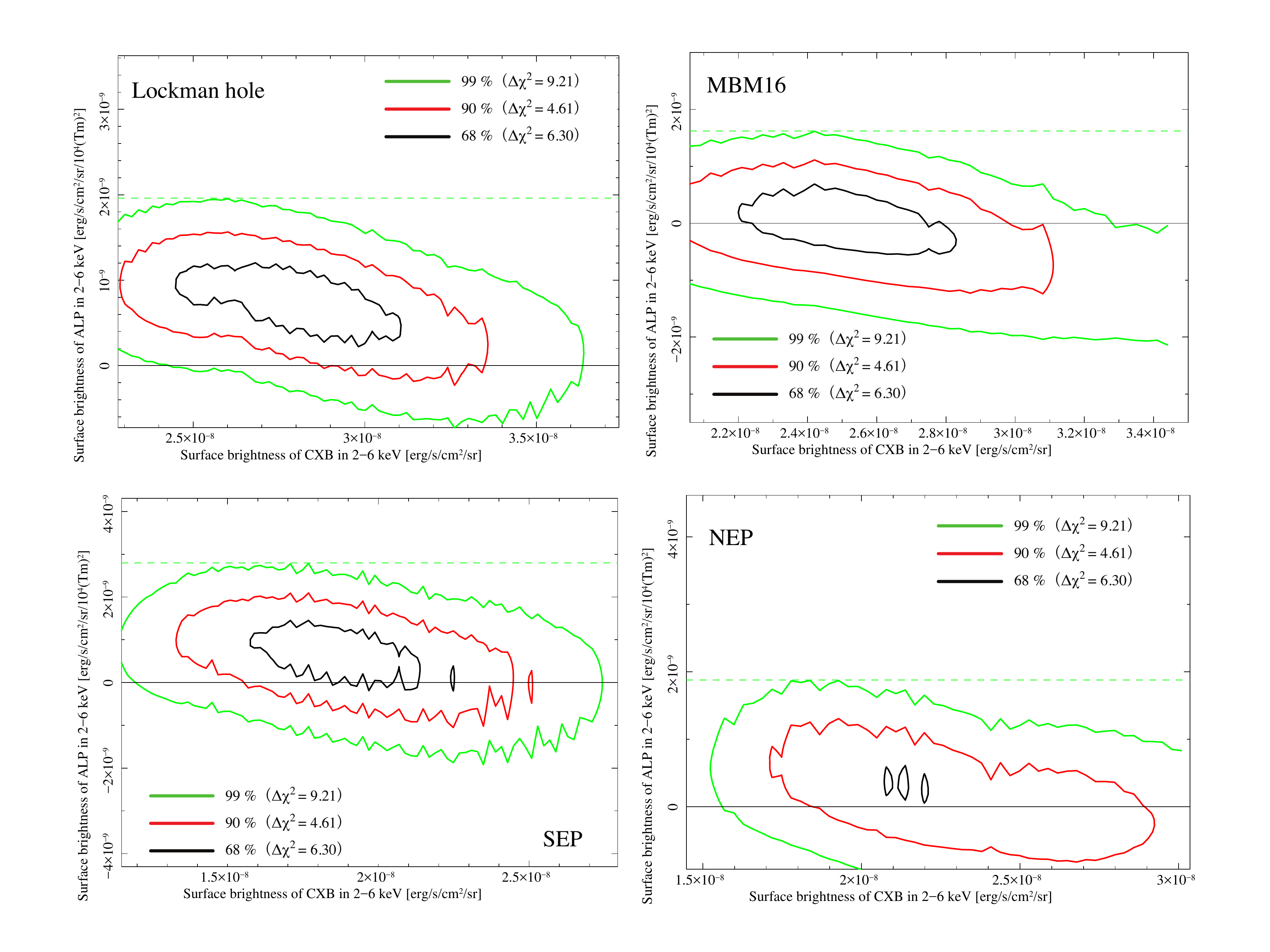}
\caption{The confidence contour between surface brightness of CXB and
 ALP calculated from the photon index
 $\Gamma_{\rm CXB}$, $\Gamma_{\rm ALP}$ and normalization
 $S_{\rm CXB}$, $S_{\rm ALP}$ as shown in Table \ref{fit_result}
 obtained for Lockman hole, MBM16, SEP, and NEP observations,
 where the NXB normalization parameters were allowed to vary.
 3 confidence levels: 68\% (black), 90\% (red) and
 99\% (green).
 Dashed line: 99\% upper limit for ALP surface brightness.}
\label{155247_29May18}
\end{center}
\end{figure}

In step \ref{164529_28May18}, we added the ALP component whose surface
brightness is proportional to $\left( B_\perp L \right)^2$. In
usual spectral fitting like Steps
\ref{162401_28May18}--\ref{164512_28May18}, we used the spectra after
subtraction of the estimated background. 
Here, we treated the NXB as one of the input models with a
normalization factor, which can be variable in each observation ID, CCD,
and $\left( B_\perp L \right)^2$ class.
In contrast, the parameters for celestial emission and the normalization
of the ALP component at a fixed $\left( B_\perp L \right)^2=10^{4}$
T$^{2}$m$^{2}$ are common for the same FoV. 
The final fitting results are summarized in Table \ref{fit_result}.
\rytrv{We assumed the flux of the ALP both in negative and positive
mathematically to evaluate proper error ranges, as shown in Figure
\ref{155247_29May18}.}

\begin{table}[htbp!]
\begin{center}
\begin{threeparttable}
\caption{Summary of best-fit parameters of the XDB + ALP + NXB model by
 spectral fitting in Lockmann Hole, MBM16, SEP, and NEP observation with
 $\left( B_\perp L \right)^2$ sorted.}
\label{fit_result}
\begin{tabular}{l l c c c c} \hline \hline
Model & Parameter & Lockman hole & MBM16 & SEP & NEP \\ \hline
Num of Obs.ID & ~ & 11 & 6 & 4 & 4 \\
\multicolumn{2}{l}{Num of $\left( B_\perp L \right)^2$
 classification$^{\ast}$} & 3 & 2 & 2 & 3 \\
Absorption & $N_{\rm H} ~ [10^{20}~{\rm cm^{-2}}]$
 & 0.58(fix) & 16.90(fix) & 4.72(fix) & 3.92(fix) \\
LHB+SWCX & $kT$ [keV] & -  & -  & 0.1(fix) & - \\
 & ${\rm Norm^{\dagger}}$ & 0(fix)& 0(fix) & $33.4^{+88.9}_{-33.4}$ &
		     0(fix) \\
MWH  & ${kT}_1$ [keV] & $0.14^{+0.08}_{-0.09}$ &
	     $0.32^{+0.24}_{-0.23}$ & - & $0.21^{+0.18}_{-0.13}$ \\
~ & ${\rm Norm}_1^{\dagger}$ & $28^{+622}_{-20}$ & $2.1^{+23.0}_{-1.3}$ &
		 0(fix) & $4.1^{+13.5}_{-3.9}$ \\
HTC & ${kT}_2$ [keV] & $0.61^{\P}$ & - & $0.66^{+0.08}_{-0.06}$ & $0.69$ \\
~ & ${\rm Norm}_2^{\dagger}$ & $0.5^{+0.4}_{-0.5} {}^{\parallel}$ & 0(fix) &
		 $2.1^{+0.5}_{-0.6}$ & $0.3^{+0.7}_{-0.3}$ \\
CXB & $\Gamma_{\rm CXB}$ & $-1.42^{+0.13}_{-0.13}$ & $-1.33^{+0.16}_{-0.16}$ &
		 $-1.50^{+0.23}_{-0.25}$ & $-1.53^{+0.18}_{-0.18}$ \\
~ & $S_{\rm CXB}^{\ddagger}$ & $7.7^{+0.6}_{-0.6}$ & $6.5^{+1.1}_{-1.0}$ &
		 $5.6^{+0.9}_{-0.8}$ & $6.9^{+0.7}_{-0.8}$ \\
ALP & $\Gamma_{\rm ALP}$ & +0.5(fix) & +0.5(fix) & +0.5(fix) & +0.5(fix) \\
~ & $S_{\rm ALP}^{\S}$ & $0.012^{+0.015}_{-0.016}$ &
	     $0.005^{+0.022}_{-0.024}$ & $0.011^{+0.027}_{-0.030}$ &
		     $0.012^{+0.020}_{-0.022}$ \\
$C/{\rm dof}$(dof) & ~ & 1.11(2865) & 1.03(1505)
 & 0.99(1000) & 1.10(1503) \\ \hline
\end{tabular}
\begin{tablenotes}[flushlef]
\begin{footnotesize}
\item All errors indicatte 90\% confidence level.
\item[$\ast$] See classification shown in Figure \ref{153849_24Jun16}.
\item[$\dagger$] The emission measure of CIE plasma integrated over the
 line-of-sight for SWCX+LHB, MWH 
 (the normalization of {\it apec} model):
 $(1/4\pi)\int n_{\rm e} n_{\rm H}ds$ in unit of
 $10^{14}~{\rm cm^{-5}~sr^{-1}}$.
\item[$\ddagger$] The surface brightness of the CXB
 (the normalization of a power-law  model): in unit of photons
 cm$^{-2}$sec$^{-1}$keV$^{-1}$str$^{-1}$ at 1 keV.]
\item[$\S$] The surface brightness of the ALP
 (the normalization of a power-law  model): in unit of photons
 cm$^{-2}$sec$^{-1}$keV$^{-1}$str$^{-1}$  at 1 keV and
 $10^4$ T$^2$ m$^2$.
\item[$\P$] Parameter pegged at fitting limit: 0.
\item[$\parallel$] Because the normalization of {\it apec} allows 0 within the
 error range, the temperature is not determined.
\end{footnotesize}
\end{tablenotes}
\end{threeparttable}
\end{center}
\end{table}

\begin{figure}[htbp!]
\begin{center}
\includegraphics[scale=0.40,bb=0 0 640 480]{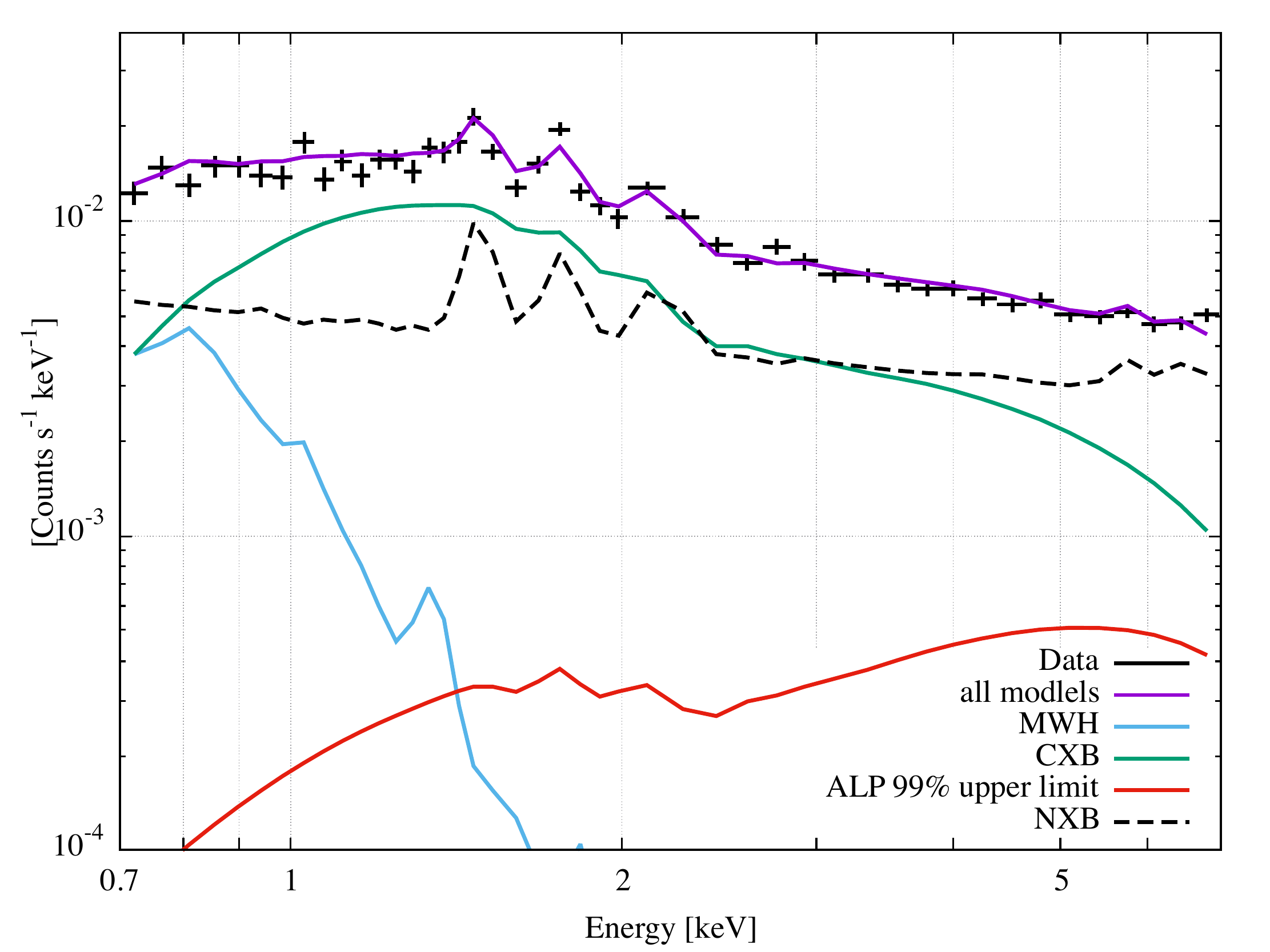}
\caption{\ryrv{Accumulated spectrum used in the spectral fit for MBM16
 direction.
 The spectrum is the sum of all Obs.ID, $(B_\perp L)^2$
 classes, and XISs. A response is weighted by the number of photons, and
 the NXB model is weighted by the exposure time after applying the
 normalization constant. Note that actual fitting was done with a set of
 energy spectra simultaneously, and no residuals are shown.}}
\label{180039_14Oct19}
\end{center}
\end{figure}

In all four observational directions, the surface brightness for the ALP
components is consistent with 0 within a 90\% confidence level.
We also checked that the normalizations of the NXB model were within
$\pm$ 40\%, or the fluctuation studied in section 3.2. 
We made contour plots by
surface brightness of ALP and CXB components between 2--6 keV, as shown
in Figure \ref{155247_29May18}.
As the surface brightness varies with the index and normalization of an
assumed power-law part, the contour is not smooth, owing to the steps in
the parameter search.
The limit obtained from the MBM16 observation 
is the lowest among these four fields and gives the tightest upper limit
on the ALP flux: $1.6 \times10^{-9}$ ergs s$^{-1}$cm$^{-2}$sr$^{-1}$
normalized at $10^{4}$ T$^{2}$m$^{2}$.
\ryrv{An accumulated spectrum of all fitted data is shown in 
Figure \ref{180039_14Oct19} with averaged XDB and NXB model and the
obtained upper limit for ALP.}
In Table \ref{upper_limit}, we also tabulated the center values
of the CXB surface brightness and the upper limits of the ratio of ALP
emission hidden in the CXB.

\begin{figure}[htbp!]
\begin{center}
\includegraphics[scale=0.30,bb=0 0 1024 768]{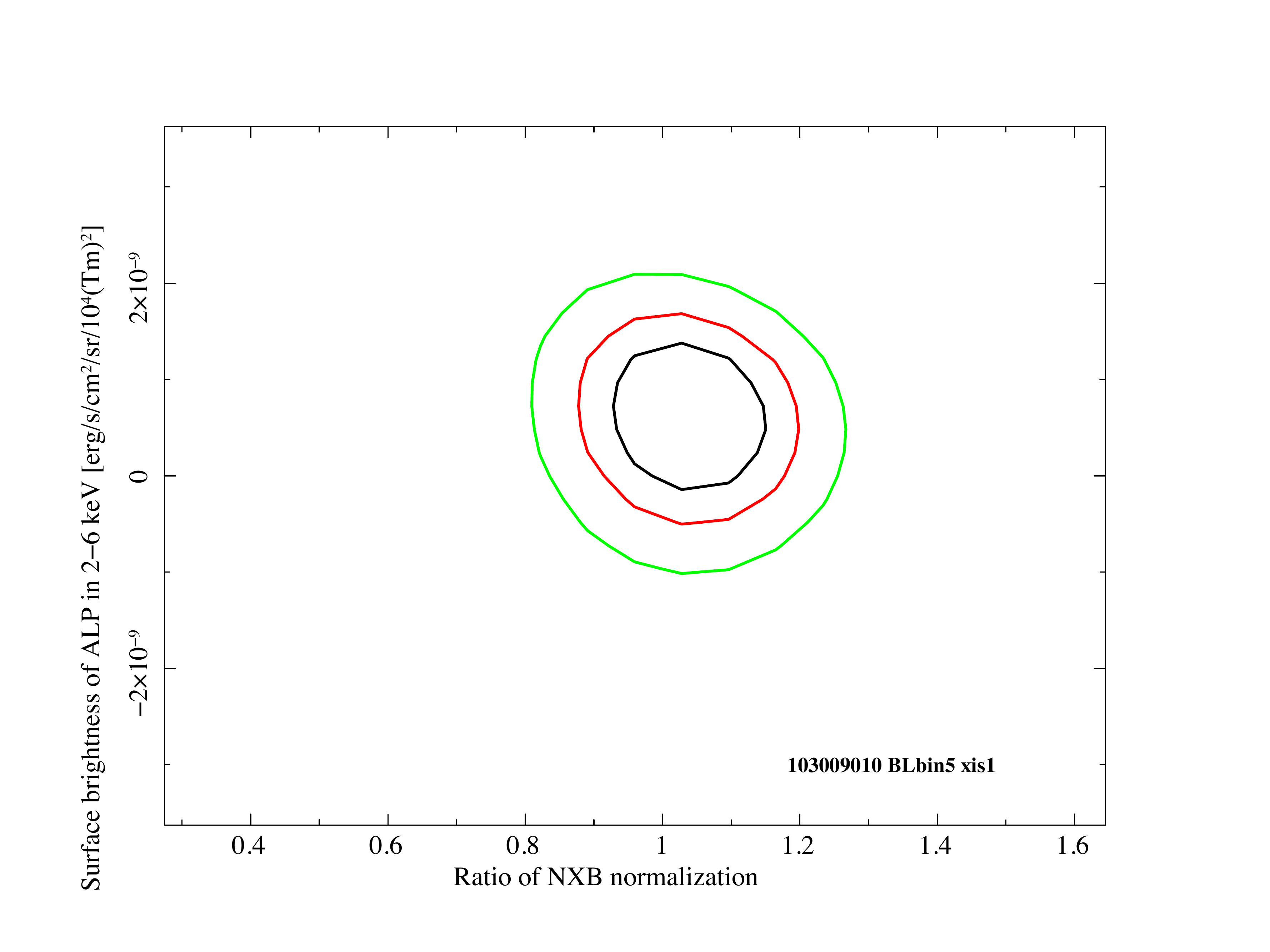}
\caption{\rytrv{The confidence contour between ratio of NXB
 normalization and surface brightness of ALP calculated from the photon
 index $\Gamma_{\rm ALP}$ and normalization $S_{\rm ALP}$ as shown in
 Table \ref{fit_result} obtained for Lockman hole at one Obs. ID, one
 $(B_\perp L)^2$ class, and one XIS.
 3 confidence levels: 68\% (black), 90\% (red) and 99\% (green).}}
\label{152346_30Nov19}
\end{center}
\end{figure}

\rytrv{To show the degeneracy among ALP and NXB normalization, we made a contour
plot with the Lockman hole observation at one Obs. ID, one BL class,
and one XIS, as shown in Figure \ref{152346_30Nov19}.
In the case of the Lockman hole, we have independently determined NXB
normalization for all 11Obs. ID, 3 BL classes, and 3 XISs.}

\begin{table}[htbp!]
\begin{center}
\begin{threeparttable}
\caption{Summary of upper limit at 99\% confidence level of surface
 brightness for ALP origin emissions as shown by dashed line in Figure
 \ref{155247_29May18}.}
\label{upper_limit}
\begin{tabular}{l c c c} \hline \hline
Field Name & 99\% UL for ALP & best-fit CXB & 99\% UL for CXB
 ratio \\
~ & surface brightness${}{\ast}$ & surface brightness${}^{\dagger}$ &
[\%] \\ \hline
Lockman hole & 2.0 & 28.3 & 7.1\\
MBM16 & 1.6 & 26.9 & 5.9 \\
SEP & 2.8 & 18.6 & 15.1 \\
NEP & 1.9 & 22.0 & 8.6 \\
 \hline
\end{tabular}
\begin{tablenotes}[flushlef]
\begin{footnotesize}
\item[$\ast$] In unit of $10^{-9}~{\rm ergs~s^{-1}~cm^{-2}~sr^{-1}}$ at
 $10^4~{\rm T^2~m^2}$ in 2--6 keV band.
\item[$\dagger$] In unit of $10^{-9}~{\rm ergs~s^{-1}~cm^{-2}~sr^{-1}}$
 in 2--6 keV band.
\end{footnotesize}
\end{tablenotes}
\end{threeparttable}
\end{center}
\end{table}

\section{Discussion and Conclusions}
\label{sec:discussion}
We assumed that the cosmologically distributed ALPs would make a
power-law with a photon index of +0.5 ($dN/dE$ $\propto$ $E^{+0.5}$)
emission by the Earth's magnetosphere in proportion to the integrated
$\left( B_\perp L \right)^2$ in the FoV, and analyzed the data with
{\it Suzaku} for four different directions.
We did not detect any possible continuous emission from ALPs reported by
previous similar studies \cite{Fraser2014}.
We obtained the 99\% upper limit of the X-ray surface brightness and
flux originating from ALPs in the 2.0-6.0 keV energy range as
1.6 $\times$ $10^{-9}$ ergs s$^{-1}$ cm$^{-2}$ sr$^{-1}$, at
$\left( B_\perp L \right)^2$ = $10^4$ T$^{2}$ m$^{2}$, as shown in Table
\ref{upper_limit}.
It corresponds to 6--15\% of the apparent CXB surface
brightness in the 2--6 keV band, and is consistent with the idea that
80--90\% of the CXB in the 2--8 keV band are resolved into point sources
\cite{Cappelluitti2017, Luo2017}.
%\cite{Brandt2015}.
In other words, it could not be denied that 10--20\% of the unresolved
CXB could originate from the ALP converted to X-ray by the Earth
atmosphere at {\it Suzaku} orbit.

\begin{figure}[htbp!]
\begin{center}
\includegraphics[scale=0.6,bb=0 0 640 480]{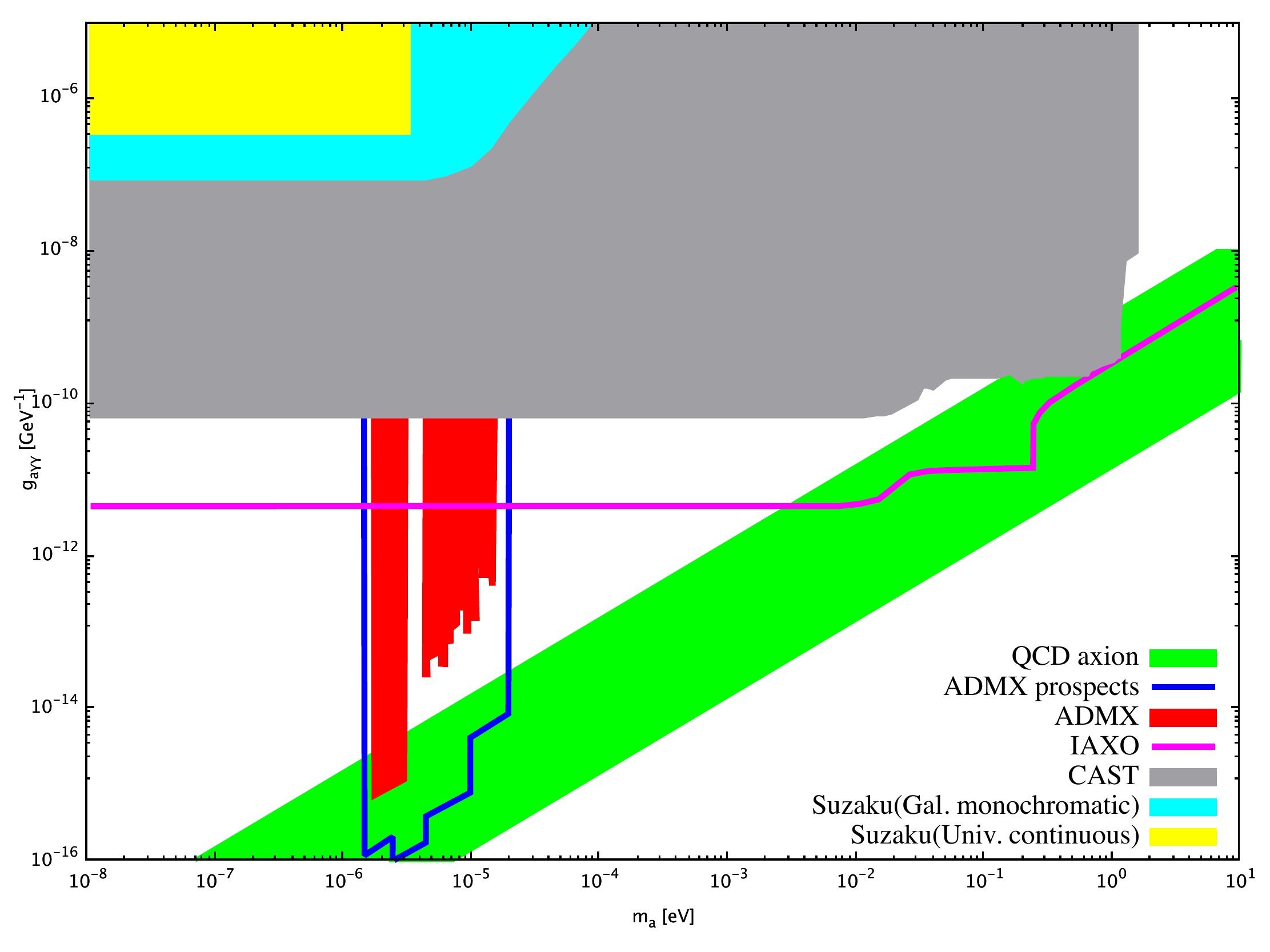}
\caption{ALP parameters constrains in this paper in Universal continuous
 ALP (cyan) and Galactic monochromatic ALP (yellow) see details in text. 
 Limits of other experiments are taken from \cite{Carosi2013,
 Anastassopoulos2017}}
\label{122724_31May18}
\end{center}
\end{figure}

If we assume the dark matter density and decay rate, we can limit the
ALP-photon coupling constant. By combining Equations \eqref{eq:con},
\eqref{124611_31May18} and \eqref{con_g10},
the ALP-photon coupling constant, $g_{a \gamma \gamma}$, was
constrained in the ALP mass range of
$m_{a}$ $<$ $\sqrt{2\pi E_a \slash L}$ $\sim$ 3.3 $\times$ 10$^{-6}$ eV
to be
\begin{eqnarray}
 g_{a \gamma \gamma} < 3.3 \times 10^{-7}~{\rm GeV^{-1}} ~
 \left( \frac{m_{\phi}}{10~{\rm keV}} \right)^{5/4}
 \left( \frac{\tau_{\phi}}{4.32\times10^{17}~{\rm ~s}} \right)^{1/2}
 \left( \frac{B_\perp L}{100~{\rm T~m}} \right)^{-1} \nonumber \\
 \left( \frac{\rho_{\phi}}{1.25~{\rm keV~cm^{-3}}} \right)^{-1/2}
 \left( \frac{H_0}{67.8~{\rm ~km~s^{-1}~Mpc^{-1}}} \right)^{-1/2}
 \left( \frac{f}{1.92} \right)^{-1/2},
\end{eqnarray}
as shown in Figure \ref{122724_31May18}.
Here, we assume a standard cosmology model with a mass density and a
Hubble constant $H_0$.
The decay rate of the dark matter to the ALPs
$\Gamma_{\phi \rightarrow 2a}$ $=$ $1/\tau_\phi$ $<$ $1/t_0$,
where $t_0$ is the Hubble time.
The factor $f$ is defined by the Equation \eqref{124611_31May18}.
\ryrv{Note that we neglect the reduction and anisotropy of the ALP flux due to
interstellar and intergalactic magnetic fields.}

For the line emission search in the X-ray band, Sekiya et al. (2016) 
collected the longest exposure of 12 Msec from 10
years of {\it Suzaku} archival data, and obtained a 3$\sigma$ upper
limit for a narrow line emission between 1 and 7 keV to be 0.021 photons
s$^{-1}$ cm$^{-2}$ sr$^{-1}$ \cite{Sekiya2016}.
The ALP-photon conversion rate, $P_{a \rightarrow \gamma}$ $\propto$
$\left( B_\perp L\right)^2$, was also computed by using IGRF-12
model every 60 seconds, and the averaged value was obtained to be
$\left( B_\perp L\right)$ = 140 T m.
It is larger than the value of 84 T m by CAST \cite{Andriamonje2007}.
This value gives the upper limits of
\begin{equation}
 I_{a,{\rm line}} \cdot (\frac{g_{a\gamma\gamma}}{10^{-10}{\rm
  GeV}^{-1}})^2 < 4.4 \times 10^{14}
 ~{\rm axions ~s^{-1} ~cm^{-2} ~sr^{-1}} \quad
 {\rm in~the~1.0-7.0~keV~band.}
\label{175448_10Nov17}
\end{equation}
A $g_{a \gamma \gamma}$ can be also constrained as 
\begin{equation}
g_{a \gamma \gamma} < 8.4 \times 10^{-8}~{\rm GeV^{-1}}
\left( \frac{B_{\perp} L}{140 {\rm ~T~m}} \right)^{-1}
\left( \frac{\tau_{\phi}}{4.32\times10^{17} {\rm ~s}} \right)^{1/2}
\left( \frac{S_{\phi}}{50 {\rm ~M_{\odot}pc^{-2}}} \right)^{-1/2},
\end{equation}
when the ALP density is connected with dark matter density around our galaxy.
It is also shown in Figure \ref{122724_31May18} as galactic monochromatic ALP.
In the plot, we consider the oscillation effect by Equation \eqref{163650_19Feb16}.
This restriction of a physical parameter of ALPs
is less  strict than other experiments (e.g. CAST, ADMX),
which assume a different axion and ALP model than this research.
Nevertheless, it is important to note that
we found these restrictions by using a new independent method from X-ray
observations.

%\appendix
%\section{Some title}
%sPlease always give a title also for appendices.

\acknowledgments
This work was partially supported by JSPS KAKENHI Grant Numbers
26220703 and 14J11023.
We thank Prof. M. Kawasaki and Prof. M. Teshima for valuable comments,
and Dr. N. Sekiya for using his data and suggestions.
We would like to thank Editage (www.editage.com) for English language
editing.


\begin{thebibliography}{99}
	
%\bibitem{Brandt2015}
%W. N. Brandt, and D. M. Alexander, \emph{Cosmic X-ray surveys of distant
%	active galaxies}, \emph{Astron. Astrophys. Rev.} {\bf vol 23}
%	(2015) 1.
	
\bibitem{Peccei1977}
R. D. Peccei, and Helen R. Quinn, \emph{{\it CP} Conservation in the
	Presence of Pseudoparticles}, \emph{Phys. Rev. Lett.} {\bf vol
	38} (1977) 1440--1443.

\bibitem{Weinberg1978}
S. Weinberg, \emph{A New Light Boson?}, \emph{Phys. Rev. Lett.} {\bf vol
	40} (1978) 223--226.

\bibitem{Cicoli2014}
M. Cicoli, J. P. Conlon, M. C. D. Marsh, and M. Rummel, \emph{3.55 keV
	photon line and its morphology from a 3.55 keV axionlike
	particle line}, \emph{Phys. Rev. D} {\bf vol 90} (2014) 023540.

\bibitem{Conlon2014}
J. P. Conlon, and F. V. Day, \emph{3.55 keV photon lines from axion to
	photon conversion in the Milky Way and
	M31}, \emph{J. Cosmol. Astropart. Phys.} {\bf vol 2014} (2014)
	033--033.

\bibitem{Conlon2015a}
J. P. Conlon, and A. Powell, \emph{A 3.55 keV line from DM $\rightarrow$
	a $\rightarrow$ $\gamma$\: predictions for cool-core and
	non-cool-core clusters}, \emph{J. Cosmol. Astropart. Phys.} 
	{\bf vol 2015} (2015) 019--019.

\bibitem{Mitsuda2007}
K. Mitsuda {\it et al.}, \emph{The X-Ray Observatory Suzaku},
	\emph{Publ. Astron. Soc. JPN} {\bf vol 59} (2007) S1--S7.

\bibitem{Fraser2014}
G. W. Fraser, A. M. Read, S. Sembay, J. A. Carter, and E. Schyns,
	\emph{Potential solar axion signatures in X-ray observations
	with the XMM-Newton observatory}, \emph{Mon. Notices Royal
	Astron. Soc.} {\bf vol 445} (2014) 2146--2168.

\bibitem{Davoudiasl2006}
H. Davoudiasl, and P. Huber, \emph{Detecting Solar Axions Using Earth's
	Magnetic Field}, \emph{Phys. Rev. Lett.}, {\bf vol 97} (2006)
	141302.

\bibitem{Davoudiasl2008}
H. Davoudiasl, and P. Huber, \emph{A feasibility study for measuring
	geomagnetic conversion of solar axions to x-rays in low Earth
	orbits}, \emph{J. Cosmol. Astropart. Phys.} {\bf vol 2008} (2008)
	026.

\bibitem{Roncadelli2015}
M. Roncadelli, and F. Tavecchio, \emph{No axions from the Sun},
	\emph{Mon. Notices Royal Astron. Lett.} {\bf vol 450} (2015)
	L26--L28.

\bibitem{Sekiya2016}
N. Sekiya, N. Y. Yamasaki, and K. Mitsuda, \emph{A search for a keV
	signature of radiatively decaying dark matter with Suzaku XIS
	observations of the X-ray diffuse background},
	\emph{Publ. Astron. Soc. JPN} {\bf vol 68} (2016) S31.

\bibitem{Kawasaki1997}
M. Kawasaki and T. Yanagida, \emph{Constraint on cosmic density of the
	string moduli field in gauge-mediated supersymmetry-breaking
	theories}, \emph{Phys. Lett. B} {\bf vol 399} (1997) 45--48.

\bibitem{Asaka1998}
T. Asaka, J. Hashiba, M. Kawasaki, and T. Yanagida, \emph{Spectrum of
	background x-rays from moduli dark matter}, \emph{Phys. Rev. D}
	{\bf vol 58} (1998) 023507.

\bibitem{VanBibber1989}
K. van Bibber, P. M. McIntyre, D. E. Morris, D. E. and G. C. Raffelt,
	\emph{Design for a practical laboratory detector for solar
	axions}, \emph{Phys. Rev. D}, {\bf vol 39} (1989) 2089--2099.

\bibitem{Thebault2015}
E. Th\'ebault, {\it et a;.},   \emph{International
	Geomagnetic Reference Field: the 12th generation}, \emph{Earth
	Planets Space} {\bf vol 67} (2015) 79.

\bibitem{Koyama2007}
K. Koyama, {\it et al.},  \emph{X-Ray Imaging
	Spectrometer (XIS) on Board Suzaku},
	\emph{Publ. Astron. Soc. Jpn} {\bf vol 59} (2007) S23--S33.
	
\bibitem{Masui2009}
K. Masui, K. Mitsuda, N. Y. Yamasaki, Y. Takei, S. Kimura, T. Yoshino,
	and D. McCammon, \emph{The Nature of Unresolved Soft X-Ray
	Emission from the Galactic Disk}, \emph{Publ. Astron. Soc. Jpn}
	{\bf vol 61} (2009) S115--S122.

\bibitem{Yamaguchi2006}
H. Yamaguchi, H. Nakajima, K. Koyama, T. G. Tsuru, H. Matsumoto,
	N. Tawa, H. Tsunemi, K. Hayashida, K. Torii, M. Namiki,
	H. Katayama, T. Dotani, M. Ozaki, H. Murakami, and E. Miller,
	\emph{The background properties of Suzaku/XIS},
	\emph{Proc. SPIE} {\bf vol 6266} (2006) 626642.
	
\bibitem{Murakami2006}
H. Murakami, M. Kitsunezuka, M. Ozaki, T. Dotani, and T. Anada,
	\emph{Origins of the instrumental background of the x-ray CCD
	camera in space studied with Monte Carlo simulation},
	\emph{Proc. SPIE} {\bf vol 6266} (2006) 62662Y.

\bibitem{Tawa2008}
N. Tawa, K. Hayashida, M. Nagai, H. Nakamoto, H. Tsunemi, H. Yamaguchi,
	Y. Ishisaki, E. D. Miller, T. Mizuno, T. Dotani, M. Ozaki, and
	H. Katayama, \emph{Reproducibility of Non-X-Ray Background for
	the X-Ray Imaging Spectrometer aboard Suzaku},
	\emph{Publ. Astron. Soc. Jpn} {\bf vol 60} (2008) S53--S76.
	
\bibitem{Kokubun2007}
M. Kokubun, {\it et al.}, \emph{In-Orbit Performance of the Hard X-Ray 
Detector on Board Suzaku},
\emph{Publ. Astron. Soc. Jpn} {\bf vol 59} (2007) S11--S24.

\bibitem{Yoshino2009}
T. Yoshino, K. Mitsuda, N. Y. Yamasaki, Y. Takei, T. Hagihara, K. Masui,
	M. Bauer, D. McCammon, R. Fujimoto, Q. D. Wang, Y, Yao,
	\emph{Energy Spectra of the Soft X-Ray Diffuse Emission in
	Fourteen Fields Observed with Suzaku},
	\emph{Publ. Astron. Soc. Jpn}, {\bf vol 61} (2009) 805--823.

\bibitem{Nakashima2018}
S. Nakashima, Y. Inoue, N. Yamasaki, Y. Sofue, J. Kataoka, T. Totani, 
\emph{Spatial Distribution of the Milky Way Hot Gaseous Halo Constrained
	by Suzaku X-Ray Observations},
\emph{Astrophys. J.},  {\bf vol 862} (2018) 34

\bibitem{Ishisaki2007}
Y. Ishisaki, Y. Maeda, R. Fujimoto, M. Ozaki, K. Ebisawa, T. Takahashi,
	Y. Ueda, Y. Ogasaka, A. Ptak, K. Mukai, K. Hamaguchi,
	M. Hirayama, T. Kotani, H. Kubo, R. Shibata, M. Ebara,
	A. Furuzawa, R. Iizuka, H. Inoue, H. Mori, S. Okada,
	Y. Yokoyama, H. Matsumoto, H. Nakajima, H. Yamaguchi,
	N. Anabuki, N. Tawa, M. Nagai, S. Katsuda, K. Hayashida,
	A. Bamba, E. D. Miller, K. Sato, and N. Y. Yamasaki, \emph{Monte
	Carlo Simulator and Ancillary Response Generator of Suzaku
	XRT/XIS System for Spatially Extended Source Analysis},
	\emph{Publ. Astron. Soc. Jpn} {\bf vol 59} (2007) S113--S132.

\bibitem{Cash1979}
W. Cash, \emph{Parameter estimation in astronomy through application of
	the likelihood ratio}, \emph{Astrophys. J.} {\bf vol 228} (1979)
	939.
	
\bibitem{Smith2001}
R. K. Smith and N. S. Brickhouse, \emph{Collisional Plasma Models with
	APEC/APED: Emission-Line Diagnostics of Hydrogen-like and
	Helium-like Ions}, \emph{Astrophys. J. Lett.}, {\bf vol 556},
	(2001) L91--95

\bibitem{Yoshitake2013}
H. Yoshitake, K. Sakai, K. Mitsuda, N. Y. Yamasaki, Y. Takei, R. Yamamoto, 
\emph{Long-Term Variability of the O VII Line Intensity toward the Lockman
 Hole Observed with Suzaku from 2006 to 2011}, 
 \emph{Publ. Astron. Soc. Jpn}, {\bf vol 65} (2013) 32.

\bibitem{Anders1989}
E. Anders, and N. Grevesse, \emph{Abundances of the elements: Meteoritic
	and solar}, \emph{Geochim. Cosmochim. Acta} {\bf vol 53} (1989)
	197--214.
 
\bibitem{Kalberla2005}
P. M. W. Kalberla, W. B. Burton, D. Hartmann, E. M. Arnal, E. Bajaja, R. Morras, 
 W. G. L.  P\"{o}ppel,  \emph{The Leiden/Argentine/Bonn (LAB) Survey of
	 Galactic HI. Final data release of the combined LDS and IAR
	 surveys with improved stray-radiation corrections},
 \emph{Astron. \& Astrophys.},  {\bf vol 440} (2005) 775--782.

\bibitem{Cappelluitti2017}
N. Cappelluti, Y. Li, A. Ricarte, B. Agarwal, V. Allevato, T. T. Ananna,
	M. Ajello, F. Civano, A. Comastri, M. Elvis, A. Finoguenov,
	R. Gilli, G. Hasinger, S. Marchesi, P. Natarajan, F. Pacucci,
	E. Treister, C. M. Urry, \emph{The Chandra COSMOS Legacy Survey:
	Energy Spectrum of the Cosmic X-Ray Background and Constraints
	on Undetected Populations}, \emph{Astrophys. J.} {\bf vol 837}
	(2017) 19.

\bibitem{Luo2017}
B. Luo, W. N. Brandt, Y. Q. Xue, B. Lehmer, D. M. Alexander,
	F. E. Bauer, F. Vito, G. Yang, A. R. Basu-Zych, A. Comastri,
	R. Gilli, Q.-S. Gu, A. E. Hornschemeier, A. Koekemoer, T. Liu,
	V. Mainieri, M. Paolillo, P. Ranalli, P. Rosati,
	D. P. Schneider, O. Shemmer, I. Smail, M. Sun, P. Tozzi,
	C. Vignali, J.-X. Wang, \emph{THE CHANDRA DEEP FIELD-SOUTH
	SURVEY: 7 MS SOURCE CATALOGS}, \emph{Astrophys. J. Suppl. S.}
	{\bf vol 228} (2017) 2.
 
 \bibitem{Andriamonje2007}
S. Andriamonje, {\it et al.}, 
	\emph{An improved limit on the axion-photon coupling from the
	CAST experiment}, \emph{J. Cosmol. Astropart. Phys.} {\bf vol
	2007} (2007) 010--010.

\bibitem{Carosi2013}
G. Carosi, A. Friendland, M. Giannotti, M. J. Pivibaroff, J. Ruz, J. K. Vogel, 
\emph{Probing the axion-photon coupling: phenomenological and
	experimental perspectives. A snowmass white paper}, \emph{
	2013arXiv1309.7035C} (2013)

\bibitem{Anastassopoulos2017}
CAST collaboration: V. Anastassopoulos, {\it et al.}, 
\emph{New CAST Limit on the Axion-Photon Interaction}, \emph{Nature
	Phys.} {\bf vol 13} (2017) 584--590 



% Please avoid comments such as "For a review'', "For some examples",
% "and references therein" or move them in the text. In general,
% please leave only references in the bibliography and move all
% accessory text in footnotes.

% Also, please have only one work for each \bibitem.

\end{thebibliography}
\end{document}